\documentclass[aps,prl,groupedaddress,showpacs,twocolumn,floatfix,superscriptaddress,groupedaddress]{revtex4}
\newcommand{\bvs}{BaVS$_3$}
\usepackage{amsmath,amssymb}	
\usepackage{graphicx}

\begin{document}


\title{Ground State of the Quasi-1D \bvs\ resolved by Resonant Magnetic X-ray Scattering}

\author{Ph.~Leininger}
\altaffiliation[]{Corresponding author: p.leininger@fkf.mpg.de}
\affiliation{Max-Planck-Institut f\"{u}r Festk\"{o}rperforschung, Heisenbergstr. 1, D-70569 Stuttgart, Germany}

\author{V. Ilakovac}\affiliation{LCP-MR, Universit\'e Pierre et Marie Curie, CNRS-UMR 7614, F-75231 Paris, France}
\affiliation{Universit\'e Cergy-Pontoise, F-95031 Cergy-Pontoise, France}

\author{Y.~Joly}\affiliation{Institut N\'{e}el, CNRS-UJF, BP 166, 38042 Grenoble Cedex 9, France}
\author{E.~Schierle}\affiliation{Helmholtz-Zentrum Berlin, Albert-Einstein-Stra{\ss}e 15, 12489 Berlin, Germany}
\author{E.~Weschke}\affiliation{Helmholtz-Zentrum Berlin, Albert-Einstein-Stra{\ss}e 15, 12489 Berlin, Germany}
\author{O.~Bunau}\affiliation{Institut N\'{e}el, CNRS-UJF, BP 166, 38042 Grenoble Cedex 9, France}
\author{H.~Berger}\affiliation{Institut de Physique de la Mati\`{e}re Complexe, EPFL, 1015 Lausanne, Switzerland}
\author{J.-P. Pouget}\affiliation{Laboratoire de Physique des Solides, Univ. Paris-Sud, CNRS, UMR 8502, F-91405 Orsay Cedex, France}
\author{P.~Foury-Leylekian}\affiliation{Laboratoire de Physique des Solides, Univ. Paris-Sud, CNRS, UMR 8502, F-91405 Orsay Cedex, France}

\date{\today}
\begin{abstract}
Resonant-magnetic x-ray scattering (RMXS) near the vanadium $L_{2,3}$-absorption edges has been used to investigate the low temperature magnetic structure of high quality \bvs\ single crystals. Below $T_N$~=~31~K, the strong resonance revealed a triple-incommensurate magnetic ordering at wave vector (0.226~0.226~$\xi$) in the hexagonal notation, with $\xi$ = 0.033. The simulations of the experimental RMXS spectra with a time-dependent density functional theory indicate an antiferromagnetic order with the spins polarized along $a$ in the monoclinic structure.
\end{abstract}
\maketitle

\bvs\ and related compounds have attracted considerable scientific interest in the last decades, mainly due to the presence of both itinerant and localized electronic states which involves a subtle interplay between charge, orbital and spin degrees of freedom~\cite{Imada1998}.
It shows a very rich phase diagram, with a metal-insulator (MI) transition at $T_{MI}$ = 69 K and another transition at around $T_X$~=~30~K that involves spin degrees of freedom.\\
However, as most of the research on \bvs\ has been done on powder samples, information about the low temperature structure is limited.  
Nuclear magnetic resonance (NMR) and nuclear quadrupole resonance measurements revealed a non magnetic ground state below $T_X$ and interpreted the asymmetric field gradient at the V sites as a consequence of orbital ordering~\cite{Nakamura1997}. Susceptibility measurements also indicated a spin pairing ground state below $T_X$~\cite{Mihaly2000}, while heat capacity measurements did not detect any transition~\cite{Imai1996}. The development of an orbital ordering between $T_{MI}$ and $T_X$ has also been suggested from dielectric function measurements~\cite{Ivek2008}.
On the other hand, powder neutron diffraction revealed very few magnetic Bragg peaks sustaining a magnetic ground state~\cite{Nakamura2000}. Supposing a ferromagnetic (FM) coupling along $c$, the structure refinement suggested an incommensurate long-range ordering with the wave vector (0.226~0.226~0)$_{H}$ (hexagonal lattice coordinates) in reciprocal lattice units (\textit{rlu}). A magnetic origin of this transition has also been supposed to explain muon spin resonance measurements [11]. Therefore the precise nature of the low temperature structure needs to be clarified.

In order to determine the exact nature of the low temperature structure, we used soft resonant magnetic x-ray scattering (RMXS), which has proved to be an excellent tool for probing magnetic ordering in $3d$ materials \cite{Abbamonte2005,Schlappa2008,Wilkins2003,Leininger2010}. The large resonant enhancement of the scattering cross section, in combination with the high brilliance of the x-ray beam provided by third generation synchrotron facilities enable the investigation of small single crystals.

In this Letter we report the direct observation of an antiferromagnetic (AFM) ground state, by RMXS at the V-$L$ edge. Below $T_N$ = 31 K, the strong resonance clearly shows an in-plane modulation, which agrees with a previous suggestion, and brings to light a small incommensurate component in the $c$-direction. Through time-dependent density functional theory (TD-DFT) calculations, we reproduce the experimental data and show that the spins are polarized along the $a$ direction in the monoclinic structure.

At room temperature the S~=~1/2 BaVS$_{3}$ (3$d^1$) system is metallic and its $P6_{3}/mmc$ structure consists of a hexagonal packing of quasi 1D-chains of face-sharing VS$_6$ octahedra along $c$, with two V atoms per unit cell ({\itshape uc})~\cite{Gardner1969}. A zigzag deformation of the V-S chains at 240 K doubles the number of (in-phase) chains in the \textit{uc} and reduces the crystal symmetry 
to an orthorhombic structure ($Cmc2_1$) ({\itshape H-O} transition). This transition is accompanied by the creation of 6 twin domains in the $ab$-plane~\cite{Fagot2005}.
Peierls instability drives the system to an insulating phase at T$_{MI}$, lowering further the symmetry to the $Im$ space group \cite{Fagot2003}. The 2$k_F$ charge density wave (CDW) is accompanied by a tetramerization of the V chains.
At $T_{MI}$ the magnetic susceptibility shows an antiferromagnetic (AFM)-like drop, 
indicating that the spin degrees of freedom are strongly affected by the MI transition 
\cite{Mihaly2000,Takano1977}. 
On the other hand, the critical wave vector 2$k_F$~=~0.5$c^{\star}$ ($c^{\star}$ stands for the reciprocal lattice vector related to $c$) indicates that only one of the two $d$ electrons per unit cell participates directly in the CDW~\cite{Fagot2003}. The effective moment deduced from the Curie-Weiss behavior of the magnetic susceptibility is 1.223 $\mu_B$, consistent with one localized spin every other V site~\cite{Fagot2003}. Band structure calculations reveal two different $t_{2g}$ active electron states at the Fermi level~\cite{Mattheiss1995,Whangbo2002,Lechermann2007}: two narrow $E_g$ bands and one dispersive $A_{1g}$ band extending along $c$ undergoing the CDW instability~\cite{Fagot2003}.

\begin{figure}[t!]
\includegraphics[width=8.5 cm]{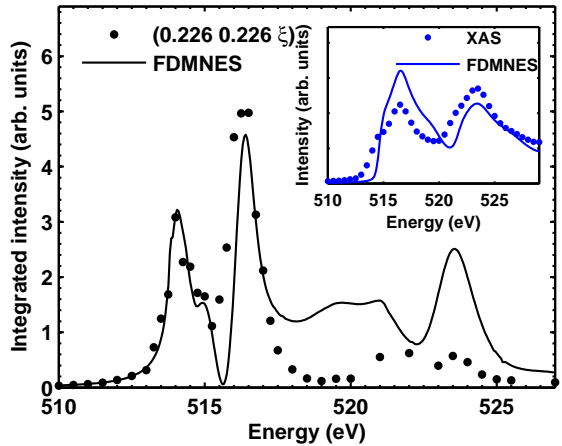}
\caption{(Color online)
Energy dependence of the magnetic reflection at (0.226 0.226 $\xi$) measured at 10 K. The solid line is the result of the {\footnotesize FDMNES} code  calculation  (see text). The inset shows the V-$L_{2,3}$ x-ray absorption spectra recorded in total fluorescence yield mode (dots) and compared with {\footnotesize FDMNES} simulation (solid line).}
\label{fluo} 
\end{figure}

The experiments were performed at the UE46-PGM1 beamline at the BESSY II synchrotron light source in Berlin, using a two-circle diffractometer with horizontal scattering geometry. The sample was mounted on a copper goniometer and attached to a continuous He-flow cryostat cooling the sample down to 10~K. The signal was recorded with a photodiode with a vertical (horizontal) acceptance of 4$^\circ$ (0.4$^\circ$) providing a momentum-resolution of 0.001 \textit{rlu}. 
The \bvs\ single crystal, grown as explained elsewhere \cite{Kuriyaki1995}, with a size of $2\times3\times0.5~mm$$^{3}$, was aligned along [110]$_{H}$, allowing a rotation about the scattering vector. The results shown here are mainly obtained on an uncleaved surface and subsequently confirmed on an in-situ cleaved surface.
\begin{figure}[b!]
\includegraphics[width=8.5 cm]{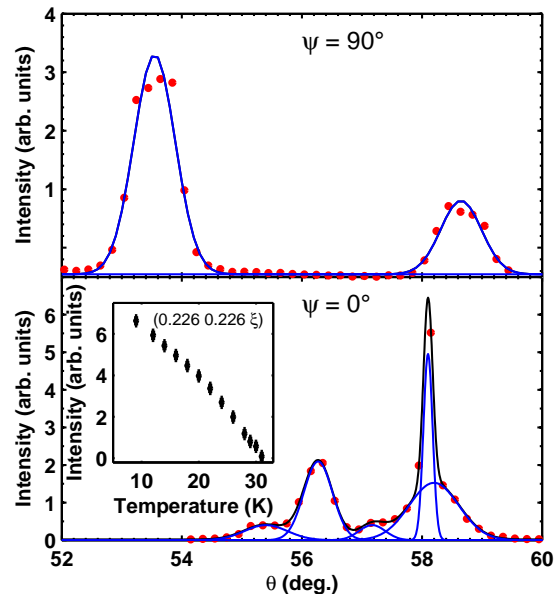}
\caption{(Color online) Rocking curve of the magnetic signal measured at $\Psi$ = 90$^\circ$ (upper panel) and at $\Psi$ = 0$^\circ$ (lower panel). The blue lines correspond to a fit with a Gaussian profile. The inset shows the temperature dependence of the (0.226 0.226 0.033) magnetic reflection measured at 10~K with 516.5 eV photon energy.}
\label{temperaturedpce}
\end{figure}

\begin{figure}[t!]
\includegraphics[width=8.5 cm]{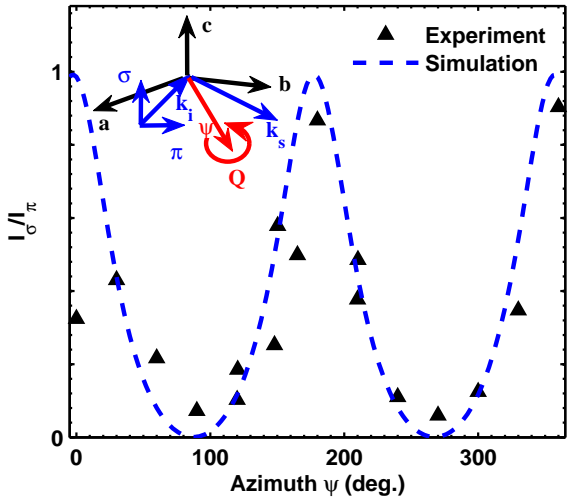}
\caption{(Color online) 
Azimuthal-angle ($\Psi$) dependence of the ratio of resonant-peak intensities measured with $\sigma$ and $\pi$ incident polarized beam (black triangles). The blue dashed line corresponds to the {\footnotesize FDMNES} simulation when the spins are polarized along a$_{Im}$ (see text). The schematic illustrates the definition of $\Psi$ and shows the scattering geometry at $\Psi$~=~0$^\circ$.}
\label{azi}
\end{figure}

At 10~K, a strong resonant enhancement of the scattered intensity is observed at wave vector (0.226$\pm 0.001$~0.226$\pm 0.001$~$\xi$)$_H$ near the V-$L_{2,3}$ absorption edges (Fig.~\ref{fluo}). This result, is the first direct observation on a single crystal, of the in-plane components of the magnetic ordering vector suggested from powder neutron diffraction~\cite{Nakamura2000}. The x-ray absorption spectra (inset Fig.~\ref{fluo}) shows two main structures with maxima at 517~eV and 524~eV, corresponding respectively to the V-$L_{3}$ (2$p_{3/2}$ $\rightarrow$ 3d) and V-$L_{2}$ (2$p_{1/2}$ $\rightarrow$ 3d) dipole transitions.
The resonant signal presents two double peak structures near 515~eV and 523~eV close to the V-$L_{3}$ and $L{_2}$ absorption edge, respectively (Fig.~\ref{fluo}). It can be attributed to different $t_{2g}$ states and related to two temperature dependent structures observed by total electron yield measurement~\cite{Vita}. This interpretation is also sustained by the calculated density of states (not shown) which shows an admixture of several $t_{2g}$ states at the Fermi level. Thereby the RMXS signal probes mainly the magnetic moment of the t$_{2g}$ states while the e$_g$ states make a negligible contribution.
The scattered intensity disappears at $T_N$ = 31$\pm$1 K (see inset Fig.~\ref{temperaturedpce})  at the same transition temperature reported in previous experiments~\cite{Mihaly2000,Nakamura2000,Higemoto2002}. Moreover, the intensity related to the square of the order parameter decreases nearly linearly when the temperature increases, as expected from a second order transition within the mean field approximation.\\
To determine precisely $\xi$, the rocking curve of the magnetic signal was measured at two different azimuthal angles $\Psi$ = 0$^\circ$ and $\Psi$ = 90$^\circ$ (Fig.~\ref{temperaturedpce}). The scattering geometry and the definition of $\Psi$ are illustrated in Fig.~\ref{azi}. Note that at $\Psi$ = 0$^\circ$ (90$^\circ$) $a$ and $b$ ($c$) are in the scattering plane.
At $\Psi = 0^\circ$ several structures related to different twin domains are clearly observed, while at $\Psi = 90^\circ$ the rocking curve shows only two well separated peaks. These two peaks, separated by $\Delta\theta$ = 4.77$^\circ$, are a fingerprint of two incommensurate magnetic reflections at (0.266~0.226~$\pm \xi$). When $\Psi = 90^\circ$ the magnetic signal from different twin domains is integrated vertically and can not explain the two structures. $\Delta\theta$ is therefore a parameter able to determine accurately the incommensurability along $c^\ast$, with a calculated value $\xi$~=~0.033 $\pm$ 0.001 in \textit{rlu}~\cite{Nakamura}.
Its small value indicates that the interactions are mainly FM in agreement with previous predictions.\\
In order to determine the spin polarization, the magnetic signal over a complete azimuthal-period has been recorded on one twin-domain. To avoid surface and sample size effects, Fig.~\ref{azi} shows the ratio $I_{\sigma}/I_{\pi}$ where $I_{\sigma}$ and $I_{\pi}$ are the intensities recorded with incoming polarization respectively perpendicular and parallel to the scattering plane. The spectra have a period $\pi$ with a maximum at $\Psi$~=~180$^\circ$ and a minimum at $\Psi$ = 90$^\circ$.
Due to the complexity of the structure and the multi-electronic phenomena occurring in the resonant scattering process, we used simulations to determine the best magnetic structure compatible with the experiments.

Simulations were performed with the {\footnotesize FDMNES} code \cite{Joly2001}. It has already been extensively used to simulate resonant elastic scattering spectra in a large variety of compounds, mainly at $K$-edges. To take into account the strong photoelectron-hole interaction occurring at the $L_{2,3}$-edges of transition metals, time-dependent density functional theory (TD-DFT), was recently incorporated in the code \cite{Bunau}. 
It was first developed by Runge \textit{et al}.~\cite{Runge1984} and later applied to simulate x-ray absorption spectra in 3$d$ transition metal compounds~\cite{Schwitalla1998}. We present here its first use in the context of RMXS.  
Calculations are fully relativistic and use a local exchange correlation Kernel (so called TD-LSDA) including the spin-orbit coupling for both the core and valence states without any approximation. To calculate the resonant atomic scattering amplitude, we use the multiple scattering frame in 5~$\AA$ radius clusters around each absorbing atom containing around 23 atoms.  
 
\begin{figure}[t]
\includegraphics{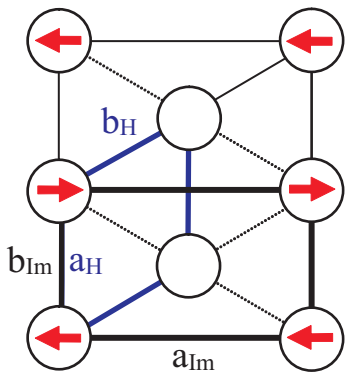}
\caption{(Color online) Magnetic structure of BaVS$_3$ in the monoclinic (black) and hexagonal (blue) lattice deduced from combined RMXS measurements and \textsl{ab inito} calculations.}
\label{structure} 
 \end{figure}

Starting from the low temperature $I_m$ monoclinic cell based on the crystal structure of Fagot {\itshape et al}.~\cite{Fagot2005}, several magnetic models have been considered. The simulations were done in the simplest 1$\times$2$\times$1 monoclinic supercell, with the spins polarized in the $ab$ plane and an AFM coupling along $b_{Im}$ (Fig.~\ref{structure}). 
In this supercell 4 layers of V atoms contains 8 inequivalent V. One layer is schematically represented in Fig.~\ref{structure}, where the 2 non-magnetic V atoms are located at the nodes of the spin wave. No spin frustration takes place and one spin S = 1/2 every two V atoms in the \textit{uc} is in agreement with Refs.~\cite{Fagot2003,Fagot2005,Fagot2006}.  
The calculations are done with the (0,-1/2,0)$_{Im}$ magnetic reflection equivalent to ($h$',-2$h$',0)$_H$, with $h$'~=~0.25. Due to the 6-fold symmetry of the hexagonal structure this reflexion is experimentally indistinguishable from the ($h$',$h$',0)$_H$. 
The calculated XAS and resonant scattering spectra reproduce the overall shape of the experimental data (Fig.~\ref{fluo}) and the excellent agreement of the peak positions indicates the correct treatment of the multi-electronic effects.

Simulations on a larger supercell, like the 1$\times$20$\times$1 with a $20b_{Im}/9$ modulation ($h$'~=~0.225 very close to $h~=~0.226$, the experimental indexation), does not change the results of the calculation. 
Adding a magnetic moment on the central V atoms does not influence the shape of the spectra but introduces a modification in the intensity.  
The calculation in the 4$\times$4$\times$1 supercell, corresponding to the ($h$',$h$',0)$_H$ reflection, leads to a spin wave along (110)$_H$ and does not affect the simulated azimuthal dependence. Moreover this cell is less reliable because the AFM ordering extends along the $a_{Im}$+$b_{Im}$ direction where the V-V distances are longer. 

The best $I_{\sigma}/I_{\pi}$ simulation is obtained when the spins on the V atoms are polarized along $a_{Im}$ (Fig.~\ref{azi}). A clear maximum (minimum) is seen at $\Psi = 180^\circ$ ($\Psi = 90^\circ$) in agreement with the experiment. 
This result is also supported by the susceptibility ($\chi$) measurements which pointed out that below $T_N$, $\chi_{ab}$ is smaller than $\chi_c$, indicating an easy axis within $ab$~\cite{Mihaly2000}.
When the spins are polarized along $b_{Im}$ the simulated azimuthal spectra is shifted by a $\pi$/2 phase with respect to the experiment, and with a magnetic moment oriented along $c_{Im}$, $I_{\sigma}/I_{\pi}$ is equal to unity.

Usually the RMXS data are analyzed with the standard formula $\left|(\hat{\epsilon}_i^{\ast}\times \hat{\epsilon}_s)\cdot \hat{m}\right|^2$~\cite{Hannon1988,Hill1996}, where $\hat{\epsilon}_i$ ($\hat{\epsilon}_s$) is the polarization of the incoming (scattered) photons and $\hat{m}$ the unit vector of the density of magnetic moment. In spite of the importance of the local symmetry in the determination of magnetic structures, not taken into account in this formula~\cite{Hannon1988,Haverkort2010}, it is valid here due to the orientation along a high symmetry direction of the magnetic moment. In fact, the simulation of the azimuthal-angle dependence with the above mentioned formula also indicates that the spins are oriented along $a_{Im}$. This attests that the multi-electronic effects do not strongly affect the sensitivity to the direction of the magnetic moment and shows that, the density of magnetic moment remains parallel to the total magnetic atomic moment.            

In summary the strong resonant enhancement observed below $T_N$ = 31~K at the V-$L_{2,3}$ edge at reflection (0.226~0.226~0.033)$_H$, reveals a triple magnetic incommensurate ordering in \bvs.
The temperature and azimuthal-angle dependence of the signal attests to the magnetic origin of the transition at $T_N$. Through simulation of the resonance with the {\footnotesize FDMNES} code in a TD-DFT multiple scattering frame, we reproduced the experimental data and established that the spins are polarized along $a_{Im}$. 
The long-range modulation along $c$ shows that the interactions are predominantly FM, indicating either a conical structure with the magnetic moment slightly deviated from the $a$-axis or a spin-density wave along~$c$.
Further theoretical investigations are needed to explain the precise origin of the out-of-plane modulation. We believe that the determination of the magnetic structure in \bvs\ will help ongoing efforts for further theoretical work which may help to understand the puzzling interplay between magnetic, charge and orbital order in low-dimensional systems. \\
{\itshape Acknowledgements}.
We are grateful to I. Zegkinoglou, M. W. Haverkort, D. Efremov, G. Jackeli, C. F. Hague and B. Keimer for discussions and helpful suggestions. We would like to thank C. Sch\"{u}{\ss}ler-Langeheine for instrumental support.

\begin{thebibliography}{29}
\expandafter\ifx\csname natexlab\endcsname\relax\def\natexlab#1{#1}\fi
\expandafter\ifx\csname bibnamefont\endcsname\relax
  \def\bibnamefont#1{#1}\fi
\expandafter\ifx\csname bibfnamefont\endcsname\relax
  \def\bibfnamefont#1{#1}\fi
\expandafter\ifx\csname citenamefont\endcsname\relax
  \def\citenamefont#1{#1}\fi
\expandafter\ifx\csname url\endcsname\relax
  \def\url#1{\texttt{#1}}\fi
\expandafter\ifx\csname urlprefix\endcsname\relax\def\urlprefix{URL }\fi
\providecommand{\bibinfo}[2]{#2}
\providecommand{\eprint}[2][]{\url{#2}}

\bibitem[{\citenamefont{Imada et~al.}(1998)\citenamefont{Imada, Fujimori, and
  Tokura}}]{Imada1998}
\bibinfo{author}{\bibfnamefont{M.}~\bibnamefont{Imada}},
  \bibinfo{author}{\bibfnamefont{A.}~\bibnamefont{Fujimori}}, \bibnamefont{and}
  \bibinfo{author}{\bibfnamefont{Y.}~\bibnamefont{Tokura}},
  \bibinfo{journal}{Rev. Mod. Phys.} \textbf{\bibinfo{volume}{70}},
  \bibinfo{pages}{1039} (\bibinfo{year}{1998}).

\bibitem[{\citenamefont{Nakamura et~al.}(1997)\citenamefont{Nakamura, Imai, and
  Shiga}}]{Nakamura1997}
\bibinfo{author}{\bibfnamefont{H.}~\bibnamefont{Nakamura}},
  \bibinfo{author}{\bibfnamefont{H.}~\bibnamefont{Imai}}, \bibnamefont{and}
  \bibinfo{author}{\bibfnamefont{M.}~\bibnamefont{Shiga}},
  \bibinfo{journal}{Phys. Rev. Lett.} \textbf{\bibinfo{volume}{79}},
  \bibinfo{pages}{3779} (\bibinfo{year}{1997}).

\bibitem[{\citenamefont{Mih\'{a}ly et~al.}(2000)\citenamefont{Mih\'{a}ly,
  K\'{e}zsm\'{a}rki, Z\'{a}mborszky, Miljak, Penc, Fazekas, Berger, and
  Forr\'{o}}}]{Mihaly2000}
\bibinfo{author}{\bibfnamefont{G.}~\bibnamefont{Mih\'{a}ly}},
  \bibinfo{author}{\bibfnamefont{I.}~\bibnamefont{K\'{e}zsm\'{a}rki}},
  \bibinfo{author}{\bibfnamefont{F.}~\bibnamefont{Z\'{a}mborszky}},
  \bibinfo{author}{\bibfnamefont{M.}~\bibnamefont{Miljak}},
  \bibinfo{author}{\bibfnamefont{K.}~\bibnamefont{Penc}},
  \bibinfo{author}{\bibfnamefont{P.}~\bibnamefont{Fazekas}},
  \bibinfo{author}{\bibfnamefont{H.}~\bibnamefont{Berger}}, \bibnamefont{and}
  \bibinfo{author}{\bibfnamefont{L.}~\bibnamefont{Forr\'{o}}},
  \bibinfo{journal}{Phys. Rev. B} \textbf{\bibinfo{volume}{61}},
  \bibinfo{pages}{R7831} (\bibinfo{year}{2000}).

\bibitem[{\citenamefont{Imai et~al.}(1996)\citenamefont{Imai, Wada, and
  Shiga}}]{Imai1996}
\bibinfo{author}{\bibfnamefont{H.}~\bibnamefont{Imai}},
  \bibinfo{author}{\bibfnamefont{H.}~\bibnamefont{Wada}}, \bibnamefont{and}
  \bibinfo{author}{\bibfnamefont{M.}~\bibnamefont{Shiga}}, \bibinfo{journal}{J.
  Phys. Soc. Jpn.} \textbf{\bibinfo{volume}{65}}, \bibinfo{pages}{3460}
  (\bibinfo{year}{1996}).

\bibitem[{\citenamefont{Ivek et~al.}(2008)\citenamefont{Ivek, Vuleti\'{c},
  Tomi\'{c}, Akrap, Berger, and Forr\'{o}}}]{Ivek2008}
\bibinfo{author}{\bibfnamefont{T.}~\bibnamefont{Ivek}},
  \bibinfo{author}{\bibfnamefont{T.}~\bibnamefont{Vuleti\'{c}}},
  \bibinfo{author}{\bibfnamefont{S.}~\bibnamefont{Tomi\'{c}}},
  \bibinfo{author}{\bibfnamefont{A.}~\bibnamefont{Akrap}},
  \bibinfo{author}{\bibfnamefont{H.}~\bibnamefont{Berger}}, \bibnamefont{and}
  \bibinfo{author}{\bibfnamefont{L.}~\bibnamefont{Forr\'{o}}},
  \bibinfo{journal}{Phys. Rev. B} \textbf{\bibinfo{volume}{78}},
  \bibinfo{pages}{035110} (\bibinfo{year}{2008}).

\bibitem[{\citenamefont{Nakamura et~al.}(2000)\citenamefont{Nakamura, Yamasaki,
  Giri, Imai, Shiga, Kojima, Nishi, Kakurai, and Metoki}}]{Nakamura2000}
\bibinfo{author}{\bibfnamefont{H.}~\bibnamefont{Nakamura}},
  \bibinfo{author}{\bibfnamefont{T.}~\bibnamefont{Yamasaki}},
  \bibinfo{author}{\bibfnamefont{S.}~\bibnamefont{Giri}},
  \bibinfo{author}{\bibfnamefont{H.}~\bibnamefont{Imai}},
  \bibinfo{author}{\bibfnamefont{M.}~\bibnamefont{Shiga}},
  \bibinfo{author}{\bibfnamefont{K.}~\bibnamefont{Kojima}},
  \bibinfo{author}{\bibfnamefont{M.}~\bibnamefont{Nishi}},
  \bibinfo{author}{\bibfnamefont{K.}~\bibnamefont{Kakurai}}, \bibnamefont{and}
  \bibinfo{author}{\bibfnamefont{N.}~\bibnamefont{Metoki}},
  \bibinfo{journal}{J. Phys. Soc. Jpn.} \textbf{\bibinfo{volume}{69}},
  \bibinfo{pages}{2763} (\bibinfo{year}{2000}).

\bibitem[{\citenamefont{Abbamonte et~al.}(2005)\citenamefont{Abbamonte, Rusydi,
  Smadici, Gu, Sawatzky, and Feng}}]{Abbamonte2005}
\bibinfo{author}{\bibfnamefont{P.}~\bibnamefont{Abbamonte}},
  \bibinfo{author}{\bibfnamefont{A.}~\bibnamefont{Rusydi}},
  \bibinfo{author}{\bibfnamefont{S.}~\bibnamefont{Smadici}},
  \bibinfo{author}{\bibfnamefont{G.~D.} \bibnamefont{Gu}},
  \bibinfo{author}{\bibfnamefont{G.~A.} \bibnamefont{Sawatzky}},
  \bibnamefont{and} \bibinfo{author}{\bibfnamefont{D.~L.} \bibnamefont{Feng}},
  \bibinfo{journal}{Nat Phys} \textbf{\bibinfo{volume}{1}},
  \bibinfo{pages}{155} (\bibinfo{year}{2005}).

\bibitem[{\citenamefont{Schlappa et~al.}(2008)\citenamefont{Schlappa,
  Schussler-Langeheine, Chang, Ott, Tanaka, Hu, Haverkort, Schierle, Weschke,
  Kaindl et~al.}}]{Schlappa2008}
\bibinfo{author}{\bibfnamefont{J.}~\bibnamefont{Schlappa}},
  \bibinfo{author}{\bibfnamefont{C.}~\bibnamefont{Schussler-Langeheine}},
  \bibinfo{author}{\bibfnamefont{C.~F.} \bibnamefont{Chang}},
  \bibinfo{author}{\bibfnamefont{H.}~\bibnamefont{Ott}},
  \bibinfo{author}{\bibfnamefont{A.}~\bibnamefont{Tanaka}},
  \bibinfo{author}{\bibfnamefont{Z.}~\bibnamefont{Hu}},
  \bibinfo{author}{\bibfnamefont{M.~W.} \bibnamefont{Haverkort}},
  \bibinfo{author}{\bibfnamefont{E.}~\bibnamefont{Schierle}},
  \bibinfo{author}{\bibfnamefont{E.}~\bibnamefont{Weschke}},
  \bibinfo{author}{\bibfnamefont{G.}~\bibnamefont{Kaindl}},
  \bibnamefont{et~al.}, \bibinfo{journal}{Phys. Rev. Lett.}
  \textbf{\bibinfo{volume}{100}}, \bibinfo{pages}{026406}
  (\bibinfo{year}{2008}).

\bibitem[{\citenamefont{Wilkins et~al.}(2003)\citenamefont{Wilkins, Hatton,
  Roper, Prabhakaran, and Boothroyd}}]{Wilkins2003}
\bibinfo{author}{\bibfnamefont{S.~B.} \bibnamefont{Wilkins}},
  \bibinfo{author}{\bibfnamefont{P.~D.} \bibnamefont{Hatton}},
  \bibinfo{author}{\bibfnamefont{M.~D.} \bibnamefont{Roper}},
  \bibinfo{author}{\bibfnamefont{D.}~\bibnamefont{Prabhakaran}},
  \bibnamefont{and} \bibinfo{author}{\bibfnamefont{A.~T.}
  \bibnamefont{Boothroyd}}, \bibinfo{journal}{Phys. Rev. Lett.}
  \textbf{\bibinfo{volume}{90}}, \bibinfo{pages}{187201}
  (\bibinfo{year}{2003}).

\bibitem[{\citenamefont{Leininger et~al.}(2010)\citenamefont{Leininger,
  Rahlenbeck, Raichle, Bohnenbuck, Maljuk, Lin, Keimer, Weschke, Schierle, Seki
  et~al.}}]{Leininger2010}
\bibinfo{author}{\bibfnamefont{P.}~\bibnamefont{Leininger}},
  \bibinfo{author}{\bibfnamefont{M.}~\bibnamefont{Rahlenbeck}},
  \bibinfo{author}{\bibfnamefont{M.}~\bibnamefont{Raichle}},
  \bibinfo{author}{\bibfnamefont{B.}~\bibnamefont{Bohnenbuck}},
  \bibinfo{author}{\bibfnamefont{A.}~\bibnamefont{Maljuk}},
  \bibinfo{author}{\bibfnamefont{C.~T.} \bibnamefont{Lin}},
  \bibinfo{author}{\bibfnamefont{B.}~\bibnamefont{Keimer}},
  \bibinfo{author}{\bibfnamefont{E.}~\bibnamefont{Weschke}},
  \bibinfo{author}{\bibfnamefont{E.}~\bibnamefont{Schierle}},
  \bibinfo{author}{\bibfnamefont{S.}~\bibnamefont{Seki}}, \bibnamefont{et~al.},
  \bibinfo{journal}{Phys. Rev. B} \textbf{\bibinfo{volume}{81}},
  \bibinfo{pages}{085111} (\bibinfo{year}{2010}).

\bibitem[{\citenamefont{Gardner et~al.}(1969)\citenamefont{Gardner, Vlasse, and
  Wold}}]{Gardner1969}
\bibinfo{author}{\bibfnamefont{R.~A.} \bibnamefont{Gardner}},
  \bibinfo{author}{\bibfnamefont{M.}~\bibnamefont{Vlasse}}, \bibnamefont{and}
  \bibinfo{author}{\bibfnamefont{A.}~\bibnamefont{Wold}},
  \bibinfo{journal}{Acta Crystallographica Section B}
  \textbf{\bibinfo{volume}{25}}, \bibinfo{pages}{781} (\bibinfo{year}{1969}).

\bibitem[{\citenamefont{Fagot et~al.}(2005)\citenamefont{Fagot,
  Foury-Leylekian, Ravy, Pouget, Anne, Popov, Lobanov, and
  Greenblatt}}]{Fagot2005}
\bibinfo{author}{\bibfnamefont{S.}~\bibnamefont{Fagot}},
  \bibinfo{author}{\bibfnamefont{P.}~\bibnamefont{Foury-Leylekian}},
  \bibinfo{author}{\bibfnamefont{S.}~\bibnamefont{Ravy}},
  \bibinfo{author}{\bibfnamefont{J.}~\bibnamefont{Pouget}},
  \bibinfo{author}{\bibfnamefont{M.}~\bibnamefont{Anne}},
  \bibinfo{author}{\bibfnamefont{G.}~\bibnamefont{Popov}},
  \bibinfo{author}{\bibfnamefont{M.}~\bibnamefont{Lobanov}}, \bibnamefont{and}
  \bibinfo{author}{\bibfnamefont{M.}~\bibnamefont{Greenblatt}},
  \bibinfo{journal}{Solid State Sci.} \textbf{\bibinfo{volume}{7}},
  \bibinfo{pages}{718} (\bibinfo{year}{2005}).

\bibitem[{\citenamefont{Fagot et~al.}(2003)\citenamefont{Fagot,
  Foury-Leylekian, Ravy, Pouget, and Berger}}]{Fagot2003}
\bibinfo{author}{\bibfnamefont{S.}~\bibnamefont{Fagot}},
  \bibinfo{author}{\bibfnamefont{P.}~\bibnamefont{Foury-Leylekian}},
  \bibinfo{author}{\bibfnamefont{S.}~\bibnamefont{Ravy}},
  \bibinfo{author}{\bibfnamefont{J.-P.} \bibnamefont{Pouget}},
  \bibnamefont{and} \bibinfo{author}{\bibfnamefont{H.}~\bibnamefont{Berger}},
  \bibinfo{journal}{Phys. Rev. Lett.} \textbf{\bibinfo{volume}{90}},
  \bibinfo{pages}{196401} (\bibinfo{year}{2003}).

\bibitem[{\citenamefont{Takano et~al.}(1977)\citenamefont{Takano, Kosugi,
  Nakanishi, Shimada, Wada, and Koizumi}}]{Takano1977}
\bibinfo{author}{\bibfnamefont{M.}~\bibnamefont{Takano}},
  \bibinfo{author}{\bibfnamefont{H.}~\bibnamefont{Kosugi}},
  \bibinfo{author}{\bibfnamefont{N.}~\bibnamefont{Nakanishi}},
  \bibinfo{author}{\bibfnamefont{M.}~\bibnamefont{Shimada}},
  \bibinfo{author}{\bibfnamefont{T.}~\bibnamefont{Wada}}, \bibnamefont{and}
  \bibinfo{author}{\bibfnamefont{M.}~\bibnamefont{Koizumi}},
  \bibinfo{journal}{J. Phys. Soc. Jpn.} \textbf{\bibinfo{volume}{43}},
  \bibinfo{pages}{1101} (\bibinfo{year}{1977}).

\bibitem[{\citenamefont{Mattheiss}(1995)}]{Mattheiss1995}
\bibinfo{author}{\bibfnamefont{L.~F.} \bibnamefont{Mattheiss}},
  \bibinfo{journal}{Solid State Commun.} \textbf{\bibinfo{volume}{93}},
  \bibinfo{pages}{791} (\bibinfo{year}{1995}).

\bibitem[{\citenamefont{Whangbo et~al.}(2002)\citenamefont{Whangbo, Koo, Dai,
  and Villesuzanne}}]{Whangbo2002}
\bibinfo{author}{\bibfnamefont{M.~H.} \bibnamefont{Whangbo}},
  \bibinfo{author}{\bibfnamefont{H.~J.} \bibnamefont{Koo}},
  \bibinfo{author}{\bibfnamefont{D.}~\bibnamefont{Dai}}, \bibnamefont{and}
  \bibinfo{author}{\bibfnamefont{A.}~\bibnamefont{Villesuzanne}},
  \bibinfo{journal}{J. Solid State Chem.} \textbf{\bibinfo{volume}{165}},
  \bibinfo{pages}{345} (\bibinfo{year}{2002}).

\bibitem[{\citenamefont{Lechermann et~al.}(2007)\citenamefont{Lechermann,
  Biermann, and Georges}}]{Lechermann2007}
\bibinfo{author}{\bibfnamefont{F.}~\bibnamefont{Lechermann}},
  \bibinfo{author}{\bibfnamefont{S.}~\bibnamefont{Biermann}}, \bibnamefont{and}
  \bibinfo{author}{\bibfnamefont{A.}~\bibnamefont{Georges}},
  \bibinfo{journal}{Phys. Rev. B} \textbf{\bibinfo{volume}{76}},
  \bibinfo{pages}{085101} (\bibinfo{year}{2007}).

\bibitem[{\citenamefont{Kuriyaki et~al.}(1995)\citenamefont{Kuriyaki, Berger,
  Nishioka, Kawakami, Hirakawa, and L\'{e}vy}}]{Kuriyaki1995}
\bibinfo{author}{\bibfnamefont{H.}~\bibnamefont{Kuriyaki}},
  \bibinfo{author}{\bibfnamefont{H.}~\bibnamefont{Berger}},
  \bibinfo{author}{\bibfnamefont{S.}~\bibnamefont{Nishioka}},
  \bibinfo{author}{\bibfnamefont{H.}~\bibnamefont{Kawakami}},
  \bibinfo{author}{\bibfnamefont{K.}~\bibnamefont{Hirakawa}}, \bibnamefont{and}
  \bibinfo{author}{\bibfnamefont{F.~A.} \bibnamefont{L\'{e}vy}},
  \bibinfo{journal}{Synth. Met.} \textbf{\bibinfo{volume}{71}},
  \bibinfo{pages}{2049} (\bibinfo{year}{1995}).

\bibitem[{\citenamefont{Vita}()}]{Vita}
\bibinfo{author}{\bibfnamefont{V.}~\bibnamefont{Ilakovac {\it et al.}}},
  \bibinfo{note}{submitted to Phys. Rev. B}.



\bibitem[{\citenamefont{Higemoto et~al.}(2002)\citenamefont{Higemoto, Koda,
  Maruta, Nishiyama, Nakamura, Giri, and Shiga}}]{Higemoto2002}
\bibinfo{author}{\bibfnamefont{W.}~\bibnamefont{Higemoto}},
  \bibinfo{author}{\bibfnamefont{A.}~\bibnamefont{Koda}},
  \bibinfo{author}{\bibfnamefont{G.}~\bibnamefont{Maruta}},
  \bibinfo{author}{\bibfnamefont{K.}~\bibnamefont{Nishiyama}},
  \bibinfo{author}{\bibfnamefont{H.}~\bibnamefont{Nakamura}},
  \bibinfo{author}{\bibfnamefont{S.}~\bibnamefont{Giri}}, \bibnamefont{and}
  \bibinfo{author}{\bibfnamefont{M.}~\bibnamefont{Shiga}}, \bibinfo{journal}{J.
  Phys. Soc. Jpn.} \textbf{\bibinfo{volume}{71}}, \bibinfo{pages}{2361}
  (\bibinfo{year}{2002}).

\bibitem[{\citenamefont{Nakamura}()}]{Nakamura}
\bibinfo{author}{\bibfnamefont{H.}~\bibnamefont{Nakamura}},
  \bibinfo{note}{suggested in a private communication the possibility of a
  small incommensurate magnetic modulation along c.}

\bibitem[{\citenamefont{Joly}(2001)}]{Joly2001}
\bibinfo{author}{\bibfnamefont{Y.}~\bibnamefont{Joly}}, \bibinfo{journal}{Phys.
  Rev. B} \textbf{\bibinfo{volume}{63}}, \bibinfo{pages}{125120}
  (\bibinfo{year}{2001}).

\bibitem[{\citenamefont{Bunau and Joly}()}]{Bunau}
\bibinfo{author}{\bibfnamefont{O.}~\bibnamefont{Bunau}} \bibnamefont{and}
  \bibinfo{author}{\bibfnamefont{Y.}~\bibnamefont{Joly}}, \bibinfo{note}{in
  preparation}.

\bibitem[{\citenamefont{Runge and Gross}(1984)}]{Runge1984}
\bibinfo{author}{\bibfnamefont{E.}~\bibnamefont{Runge}} \bibnamefont{and}
  \bibinfo{author}{\bibfnamefont{E.~K.~U.} \bibnamefont{Gross}},
  \bibinfo{journal}{Phys. Rev. Lett.} \textbf{\bibinfo{volume}{52}},
  \bibinfo{pages}{997} (\bibinfo{year}{1984}).

\bibitem[{\citenamefont{Schwitalla and Ebert}(1998)}]{Schwitalla1998}
\bibinfo{author}{\bibfnamefont{J.}~\bibnamefont{Schwitalla}} \bibnamefont{and}
  \bibinfo{author}{\bibfnamefont{H.}~\bibnamefont{Ebert}},
  \bibinfo{journal}{Phys. Rev. Lett.} \textbf{\bibinfo{volume}{80}},
  \bibinfo{pages}{4586} (\bibinfo{year}{1998}).

\bibitem[{\citenamefont{Fagot et~al.}(2006)\citenamefont{Fagot,
  Foury-Leylekian, Ravy, Pouget, Lorenzo, Joly, Greenblatt, Lobanov, and
  Popov}}]{Fagot2006}
\bibinfo{author}{\bibfnamefont{S.}~\bibnamefont{Fagot}},
  \bibinfo{author}{\bibfnamefont{P.}~\bibnamefont{Foury-Leylekian}},
  \bibinfo{author}{\bibfnamefont{S.}~\bibnamefont{Ravy}},
  \bibinfo{author}{\bibfnamefont{J.-P.} \bibnamefont{Pouget}},
  \bibinfo{author}{\bibfnamefont{E.}~\bibnamefont{Lorenzo}},
  \bibinfo{author}{\bibfnamefont{Y.}~\bibnamefont{Joly}},
  \bibinfo{author}{\bibfnamefont{M.}~\bibnamefont{Greenblatt}},
  \bibinfo{author}{\bibfnamefont{M.~V.} \bibnamefont{Lobanov}},
  \bibnamefont{and} \bibinfo{author}{\bibfnamefont{G.}~\bibnamefont{Popov}},
  \bibinfo{journal}{Phys. Rev. B} \textbf{\bibinfo{volume}{73}},
  \bibinfo{pages}{033102} (\bibinfo{year}{2006}).

\bibitem[{\citenamefont{Hannon et~al.}(1988)\citenamefont{Hannon, Trammell,
  Blume, and Gibbs}}]{Hannon1988}
\bibinfo{author}{\bibfnamefont{J.~P.} \bibnamefont{Hannon}},
  \bibinfo{author}{\bibfnamefont{G.~T.} \bibnamefont{Trammell}},
  \bibinfo{author}{\bibfnamefont{M.}~\bibnamefont{Blume}}, \bibnamefont{and}
  \bibinfo{author}{\bibfnamefont{D.}~\bibnamefont{Gibbs}},
  \bibinfo{journal}{Phys. Rev. Lett.} \textbf{\bibinfo{volume}{61}},
  \bibinfo{pages}{1245} (\bibinfo{year}{1988}).

\bibitem[{\citenamefont{Hill and McMorrow}(1996)}]{Hill1996}
\bibinfo{author}{\bibfnamefont{J.~P.} \bibnamefont{Hill}} \bibnamefont{and}
  \bibinfo{author}{\bibfnamefont{D.~F.} \bibnamefont{McMorrow}},
  \bibinfo{journal}{Acta Crystallographica Section A}
  \textbf{\bibinfo{volume}{52}}, \bibinfo{pages}{236} (\bibinfo{year}{1996}).

\bibitem[{\citenamefont{Haverkort et~al.}(2010)\citenamefont{Haverkort,
  Hollmann, Krug, and Tanaka}}]{Haverkort2010}
\bibinfo{author}{\bibfnamefont{M.~W.} \bibnamefont{Haverkort}},
  \bibinfo{author}{\bibfnamefont{N.}~\bibnamefont{Hollmann}},
  \bibinfo{author}{\bibfnamefont{I.~P.} \bibnamefont{Krug}}, \bibnamefont{and}
  \bibinfo{author}{\bibfnamefont{A.}~\bibnamefont{Tanaka}},
  \bibinfo{journal}{Phys. Rev. B} \textbf{\bibinfo{volume}{82}},
  \bibinfo{pages}{094403} (\bibinfo{year}{2010}).

\end{thebibliography}
\bibliographystyle{apsrev}

\end{document}